\documentstyle[12pt]{ioplppt}
\begin{document}
\newcommand{\dj} {\mbox{d\raisebox{0.75ex}{\hspace*{-0.32em}-}\hspace*{-0.02em}}}
\newcommand{\Dj} {\mbox{D\raisebox{0.30ex}{\hspace*{-0.75em}-}\hspace*{ 0.42em}}}
\jl{4}
\letter{The  importance  of the nucleon-nucleon correlations for the $\eta
        \alpha$ S-wave scattering length, and the $\pi  ^0$-$\eta$  mixing
        angle in the low energy $\eta \alpha$ scattering length model}

\author{S Ceci\dag, D Hrupec\ddag\ftnote{3}{dario.hrupec@fer.hr}
        and A \v{S}varc\dag\ftnote{4}{svarc@rudjer.irb.hr}}

\address{\dag\ Ru{\dj}er Bo\v{s}kovi\'{c} Institute, Zagreb, Croatia}
\address{\ddag\ University of Zagreb, Faculty of Electrical Engineering
                and Computing, Zagreb, Croatia}

\begin{abstract}
      Using  the  new set of $dd \rightarrow \eta \alpha$ near threshold
      experimental  data,  the  estimate  of  the  importance   of   the
      nucleon-nucleon   correlations   for   the  $\eta  \alpha$  S-wave
      scattering length in the multiple scattering  theory  is  obtained
      using  the  low  energy scattering length model.  The contribution
      turns out  to  be  much  bigger  then  previously  believed.   The
      $\pi^0$-$\eta$  mixing  angle  is extracted using the experimental
      data on the $dd \rightarrow \eta  \alpha$  and  $  dd  \rightarrow
      \pi^0   \alpha$   processes.    The  model  is  dominated  by  the
      subthreshold extrapolation recipe for the $\eta \alpha$ scattering
      amplitudes.   When  the  recipe  is chosen the model is completely
      insensitive to the $\eta \alpha$ parameters for  the  subthreshold
      value  of  the  $\eta$  cm   momentum  of $p_{\eta}^2=-(0.46)^2$
      fm$^{-2}$.  Provided that the subthreshold extrapolation recipe is
      correct,  a  good  estimate of the $\pi ^0$-$\eta$ mixing angle is
      obtained  if  the  experimental  cross  sections   for   the   $dd
      \rightarrow  \pi ^0 \alpha$ reaction at the corresponding deuteron
      input energy are taken from the literature.  
\end{abstract}

\pacs{13.60.Le, 13.75.n, 25.10.+s, 25.40.Ve}
\maketitle

\nosections

\newpage
\setcounter{page}{1}
      In the calculation of the $\eta \alpha$ S-wave scattering  length,
      nucleon-nucleon    correlation    corrections   to   the   impulse
      approximation  in  the  multiple  scattering  theory   have   been
      previously     investigated     within     different    formalisms
      \cite{Eri69,Wyc93,Wil94}.  The agreement about its absolute  value
      has  not  been reached.  We have used the low energy $\eta \alpha$
      scattering length model to estimate that value on the basis of the
      experimental  data  for  the  $dd  \rightarrow  \eta  \alpha$ near
      threshold measurements \cite{Fra94,Wil97} and have shown that  the
      obtained value is significantly bigger then in \cite{Wil94}.

      In  spite  of  the  fact  that  the  $dd \rightarrow \pi^0 \alpha$
      process is forbidden by the isospin conservation,  nonzero  values
      for  the  total  cross section for that process have been reported
      \cite{Gol91}.   Unfortunately,  these   measurements   have   been
      performed  at  the  subthreshold  energy  for  the isospin allowed
      $\eta$ production.  The near-threshold measurements  for  the  $dd
      \rightarrow  \eta  \alpha$ can not, therefore, be directly used to
      extract the  $\pi^0$-$\eta$  isospin  mixing  parameter. The
      dominant  $\eta$-production  S-wave  scattering function has to be
      extrapolated below $\eta$  production  threshold  to  determine  a
      charge  symmetry  breaking $\pi^0$-$\eta$ mixing angle. It is shown
      that the extracted value of this angle almost completely depends on
      the subthreshold  extrapolation
      recipe  for  the  $\eta  \alpha$  scattering  function,  and it is
      extremely  insensitive  to  the  details  of  the  $\eta   \alpha$
      interaction.

      When  the  experimental  data  quite  near  the  $\eta$ production
      threshold with the deuteron energy varying only  several  MeV  are
      used,  the  scattering amplitude must be associated with the $\eta
      \alpha$  final  state  interaction.   Therefore,  the   scattering
      amplitude  $f_{\eta}$  of the process $dd \rightarrow \eta \alpha$
      can be written as \cite{Gol64}:
\begin{eqnarray}
      \label{eq:len}
      f_{\eta} &=& \frac{f_B}{p_{\eta} a_{\eta \alpha} \cot \delta-\imath \
                              p a_{\eta \alpha} }
\end{eqnarray}
      $a_{\eta \alpha}, \delta$ ...  the  scattering length and  the S-wave
      phase shift in the  exit  channel \\
      $p_{\eta}$ ... the  cm momentum  in  the  exit  \vspace*{1.cm} 
      channel \\
      If the usual approximation for the weak transition  to  a  channel
      with  a  strong  final-state  interaction is used \cite{Gol64} the
      function $f_B$ is a slowly varying function near $\eta$ production
      threshold.   If  near threshold expansion of the S-wave scattering
      phase shift $\delta$ is applied \cite{Tay72}:
\begin{eqnarray}
      p_{\eta} \cot \delta &=& \frac{1}{a_{\eta \alpha} }   \\ \nonumber 
\end{eqnarray}
      the  formula  Equation(\ref{eq:len})  is  transformed  to the
      following form:
\begin{eqnarray}
      \label{eq:feta}
      f_{\eta} &\approx & \frac{f_B}{1 - \imath \ p_{\eta} a_{\eta \alpha} } 
\end{eqnarray}
      Finally, the square of the absolute value of the $\eta \alpha$
      scattering amplitude is expressed in terms of the differential
      cross section as:
\begin{eqnarray}
      \label{eq:xs}
      |f_{\eta}|^2 & = & \frac{p_d}{p_{\eta}} \frac{d \sigma}{d \Omega}
\end{eqnarray}

      We suggest to parametrize the nucleon-nucleon  correlations  using
      the  multiple-scattering  theory,  as  it  has  been  done  in 
      reference \cite{Wil94}.

      In the multiple-scattering  expansion  the  $\eta  \alpha$  S-wave
      scattering  length  depends  on  the  impulse  approximation  term
      (dependent on the $\eta$-nucleon S-wave scattering length) and the
      nucleon-nucleon correlations contribution $\beta$ in the following
      way:
\begin{eqnarray}
      \label{eq:msc}
      \frac{1}{a_{\eta \alpha}} &=& \frac{1}{4 R a_{\eta N}} -
              \beta. 
\end{eqnarray}
      In reference \cite{Wil94} the $\beta$ is assumed to be real and
      $R=m_{red}(\eta  \alpha)/m_{red}(\eta  N)  \approx 1.38$.
      The subscript $_{red}$ means that the reduced masses are
      used.The assumption of reality of $\beta$ bears no physical
      meaning, and is
      used to make the model more transparent. Introducing
      the imaginary part would just bring in an additional free parameter
      having no  obvious physical interpretation.

      However,  the  inputs  to this equation have not been well defined
      until recently:  both, the numerical value of the real part of the
      $a_{\eta  N}$  (S-wave scattering length), and the nucleon-nucleon
      correlation factor have been model dependent.  The origin  of  the
      $\eta  N$ scattering length problem has been extensively discussed
      in ref.  \cite{Sva96}.  The real  part  of  the  $\eta  N$  S-wave
      scattering  length  was  reported  to  have very different values:
      0.27 \cite{Bha85} $\leq$ Real($a_{\eta N}) \leq$ 0.98 fm \cite{Ari92}.
      In all cases the imaginary part is quite well fixed by the optical
      theorem  (Im($a_{\eta  N})  \approx$  0.26   fm).    Recently,   a
      controversy  is  resolved,  and a general agreement on the size of
      the real part has been reached \cite{Wil97,Gre97,Sva98,Kul98}.  It
      is  agreed  that  it  is  definitely  bigger then 0.5 fm, and 
      close to 0.72 fm.  In this  article  we  have  used  four
      values:   Real($a_{\eta  N})=$  0.35  fm  \cite{Exa98};   0.48  fm
      \cite{Wil93};  0.55 fm \cite{Wil93a} and 0.72 fm  \cite{Sva98}  as
      an illustration of the problem.
     
      The nucleon-nucleon correlation factor $\beta$ is,  on  the  other
      hand,  theoretically  estimated  in the simple multiple-scattering
      expansion for the S-wave scattering length $a_{\eta \alpha}$.   In
      reference \cite{Wil94} it has been approximated  with $\beta=0.75
      \langle \frac{1}{x} \rangle$, x being the separation  between  two
      nucleons  in  the  $\alpha$  target,  and  $R$ is the afore defined
      ratio of the reduced masses. The factor 3/4 originates from the
      self-correlations, and
      the  $\langle  \frac{1}{x}  \rangle$  factor  is  for   simplicity
      estimated in the rigid model of the $\alpha$ particle to the value
      of  0.375  fm$^{-1}$ \cite{Wil94}. The  question  arises  whether
      such   an
      approximation  for  the nucleon-nucleon correlations is compatible
      with the value obtained from the experiment, and that is  what  we
      have  done  using  the  low energy $\eta \alpha$ scattering length
      model.

      The Argand diagram for the $\eta \alpha$ S-wave  scattering  length
      is shown in Figure 1 for several suggested values of the $\eta N$
      S-wave scattering length, and as a function of $\beta$.  The  open
      squares  connected  with the full line show the value of the $\eta
      \alpha$ scattering  length  for  the  nucleon-nucleon  correlation
      factor   $\beta=0.28$, corresponding to the estimate of reference
      \protect{\cite{Wil94}} for different values of $a_{\eta N}$.  Full
      dots  connected  with  dotted lines show the value of the $a_{\eta
      \alpha}$ for different  $a_{\eta  N}$  values  and  for  different
      values  of $\beta$ in steps of 0.05 fm$^{-1}$, and $0 < \beta < 0.6$
      fm$^{-1}$.  The meaning of the open  triangle  will  be  discussed
      later.

      We  have  decided to test which value of $\beta$ correspond to the
      experimental data in the following way:

      We  take  squares  of  the  experimental  values of the scattering
      amplitude for the $dd \rightarrow \eta \alpha$  process  from  the
      literature  \cite{Fra94,Wil97}.  We take the low energy scattering
      length model given by Equation (\ref{eq:xs}),  and  normalize  the
      value  of the function $|f_{\eta}|^2$ at the point $p_{\eta}$=0.15
      fm$^{-1}$ to the value of $|f_{\eta}|^2$=27.0  nb/sr  \cite{Wil97}
      for  different  $\beta$ values.  The results for $\beta=0.28;0.42$
      and 0.56 fm$^{-1}$ and for the  $\eta  N$  S-wave  scattering
      length   of  ref.   \cite{Sva98}  are  shown  in  Figure  2,  for
      $p_{\eta}>0$.  The agreement of the model with experiment for
      the  $\eta  N$  S-wave scattering length of ref.  \cite{Wil94} and
      $\beta$=0.28 is not shown here, but the reader is refereed to  the
      original  reference  - see.  \cite{Wil94}, Fig.1. The other part
      of the figure ($p_{\eta}  <  0$)  will  be  explained  later.   We
      conclude  that  the best agreement with experiment is obtained for
      the value $\beta=0.56$ fm$^{-1}$.  Therefore, the  nucleon-nucleon
      correlation  corrections  which reproduce the experimental numbers
      for  $dd  \rightarrow  \eta   \alpha$   are   much   bigger   then
      theoretically  estimated in reference \cite{Wil94}, and correspond
      to the $\eta \alpha$ S-wave scattering length  value  of  $a_{\eta
      \alpha}=(-2.88  +  \imath \ 0.71)$ fm.  That significantly differs
      from the value $a_{\eta \alpha}= (0.396 + \imath  \  1.43)$  which
      would  come  out as a result of a simple impulse approximation and
      from the value $a_{\eta \alpha}=( 0.06 + \imath \ 6.02)$ fm  given
      in reference \cite{Wil94}.  On the other hand, the obtained result
      quite well corresponds  to  the  value  $a_{\eta  \alpha}=(-2.2  +
      \imath  \  1.1)$  fm  of  Willis  et  al.   \cite{Wil97}  which is
      extracted by direct fitting the same experimental data set but not
      in the multiple-scattering formalism basically defined by Equation
      (\ref{eq:msc}).  That value of  $\eta  \alpha$  S-wave  scattering
      length is represented by the inverse triangle in Figure 1.

      The  $\pi ^0$-$\eta$ mixing angle is defined in the following way:
      \\ $\pi ^0$ and $\eta$ have  identical  quantum  number  with  the
      exception  of  isospin, and the physical particles are formed as a
      mixture of pure isospin states.  Then the mixing angle $\theta$ is
      given as:
\begin{eqnarray}
     | \pi ^0 \rangle &=&\ \ \cos \theta |\tilde{\pi ^0}\rangle
                     +  \sin \theta |\tilde{\eta} \rangle  \\
     | \eta  \rangle & =& -\sin \theta |\tilde{\pi ^0}\rangle
                     +  \cos \theta |\tilde{\eta} \rangle \nonumber
\end{eqnarray}
      where  $|\tilde{\pi  ^0}\rangle$  and  $|\tilde{\eta} \rangle$ are
      isospin  eigenstates. If we follow the formalism of reference
      \cite{Wil94},  the  $\pi ^0$-$\eta$ mixing angle is extracted from
      the ratio of the subthreshold extrapolation of the $dd \rightarrow
      \eta  \alpha$  cross  section  to  the cm $\eta$ momentum value of
      $p_{\eta}^2=-(0.46)^2$ fm$^{-2}$ and the measured value of the $dd
      \rightarrow   \pi   ^0   \alpha$   scattering   function   at  the
      corresponding point \cite{Gol91}.  The subthreshold  extrapolation
      of  the  $\eta  d$  amplitude is not known, and in the afore cited
      model it is defined by the assumption  that  the  $\eta$  momentum
      becomes  complex,  and the absolute value keeps the negative sign.
      Then, the mixing angle $\theta$ is extracted as:
\begin{eqnarray}
      \cos \theta &=& \frac{1}{\sqrt{1+\lambda} } \\
      \lambda&=& \frac{p_d}{p_\pi F_N} \frac{d\sigma}{d\Omega}
                     (dd \rightarrow \pi ^0 \alpha) \nonumber \\
      F_N&=& N F F^\star \nonumber \\
      F&=& \frac{a_{\eta \alpha}}{1-\imath \ a_{\eta \alpha} p_{\eta} }
      \nonumber
\end{eqnarray}
      where $p_d$ and $p_\pi$ are deuteron and pion cm  momentum  values
      at    the    subthreshold    $\eta$    production    momentum   of
      $p_{\eta}^2=-(0.46)^2$  fm$^{-2}$  and  $\lambda$  is   the   $\pi
      ^0$-$\eta$  mixing parameter.
      The measured $dd \rightarrow \eta \alpha$ cross sections  are
      "hidden" in the $a_{\eta \alpha}$ scattering length parameter.
      The $N$ is a normalization constant
      which ensures that the low energy scattering  amplitude  expansion
      is  reproducing  the  measured value of the square of the absolute
      value of the scattering amplitude of 27.0  nb/sr  for  the  $\eta$
      momentum  of 0.15 fm$^{-1}$ for the $dd \rightarrow \pi ^0 \alpha$
      process \cite{Wil97} (effectively simulating  the  $f_B$  function
      given in Equation (\ref{eq:len}).

      According   to  the  ref.   \cite{Gol91}  the  value  of  the  $dd
      \rightarrow \pi ^0  \alpha$  differential  cross  section  at  the
      deuteron  kinetic  energy of T$_d$=1.100 GeV, more specifically 20
      MeV below $\eta$ production threshold is
\begin{center}
        $\frac{d \sigma}{d \Omega} = (1.00 \pm 0.25)  \ \ \frac{\rm pb}
                         {\rm sr}$
\end{center}
      at the cm  angle of 73$^0$.  However,  this  result  has  to  be
      taken  "with the grain of salt" because of the presence of the two
      photon  (or  e$^+$e$^-$)  events  observed  in   the   experiment.
      Combined with the cuts imposed by the acceptance and the analysis,
      this continuum might simulate a "$\pi ^0$" event of  approximately
      the  right mass \cite{Wil94}.  Nevertheless, it is interesting to
      wonder, if we take the result at face value,
      how accurately can we estimate $\pi
      ^0$-$\eta$ charge symmetry breaking parameter. As the result for the
      $dd \rightarrow \pi^0 \alpha$ process is to be confirmed, we try
      to estimate how much the $\pi^0$-$\eta$ mixing parameter depends
      on the uncertainty, and not only on the statistical one given in the
      article, but as well on the systematic one (still quite opened).

      If  we  take  the  value  of  $\beta=0.56$  fm$^{-1}$ for $a_{\eta
      N}=(0.72+ \imath \ 0.26)$ fm, the  values  of  the  $\pi
      ^0$-$\eta$ mixing parameters are uniquely extracted:
\begin{eqnarray}
       \label{eq:theta}
       \theta=0.986^0      & \ \ \ \  & \lambda=0.017
\end{eqnarray}
      However,  it is very interesting to observe that the values of the
      square of the absolute value of the $\eta  \alpha$  amplitude  are
      extremely   insensitive   to   the   nucleon-nucleon  correlations
      parameter $\beta$, and henceforth to  the  overall  $\eta  \alpha$
      S-wave scattering length,see Figure 2. The value of $|f_{\eta}|^2$
      at  $p_{\eta}^2=-(0.46)^2$  fm$^{-2}$  is  almost  independent  of
      $\beta$.   The insensitivity to the $\eta \alpha$ input originates
      from  the  fact  that  within  the  low  energy  model,  only  the
      subthreshold  extrapolation  of  the  $\eta  N$  S-wave scattering
      amplitude determines the shape of the $|f_{\eta}|^2$ curve. As it can
      be seen in Figure 2 the tail is mostly insensitive to the details
      of the $\eta \alpha$ S-wave scattering length, and that is  exactly
      the domain where the $\pi ^0$-$\eta$ mixing angle determination is
      performed.  Introducing the $\eta \alpha$  effective  range  might
      change  the  afore  conclusion slightly, but as it is a completely
      unknown parameter it will not improve the predictive power of  the
      model.

      The insensitivity of the model to the $a_{\eta \alpha}$ is used to
      make a correlation of the $\pi ^0$-$\eta$  mixing  parameters  and
      the  measured $dd \rightarrow \pi ^0 \alpha$ cross section at that
      energy.  In Table 1 we show that correspondence assuming that the
      systematic error for the $dd \rightarrow \pi^0 \alpha$ measurement
      can range from 1 pb/sr to maximally 5 pb/sr. Otherwise, the signal
      would be clearly detected.  \\

\begin{table}
\caption{The experimental value of the $dd \rightarrow \pi ^0 \alpha$
           cross section as a function of different $\pi ^0$-$\eta$ mixing
         parameters at $p_{\eta}=-(0.46)^2 fm^{-1}$.}
\begin{indented}
\item[]\begin{tabular}{ccc}
\br
d$\sigma$/d$\Omega$ (pb/sr)     &     $\lambda$       &      $\theta^0$ \\
\mr
      1                         &     0.017           &      0.980      \\
      2                         &     0.024           &      1.394      \\ 
      3                         &     0.030           &      1.707      \\
      4                         &     0.034           &      1.971      \\
      5                         &     0.038           &      2.204      \\
\br
\end{tabular}
\end{indented}
\end{table}

     We can offer two general conclusions:
\begin{itemize}
      \item
      The  nucleon-nucleon  correlation  contributions  to  the  impulse
      approximation  for the calculation of the $\eta \alpha$ scattering
      length in the multiple-scattering  theory  are  much  higher  then
      previously  anticipated  \cite{Wil94} and the $\eta \alpha$ S-wave
      scattering length of $a_{\eta  \alpha}=(0.06 +  \imath  \ 6.02)$  fm
      obtained in that article should not be taken as realistic.
      \item  The precision of experimental  resolution  of  the  deuteron
      beam kinetic energy is seen as a possible problem (the lab kinetic
      energy for dd initial state goes in steps of  1  Mev  at  the  GeV
      level).   However,  we  have taken the published numbers at a face
      value and we are not discussing how reliable they are.  We analyze
      the  impact  of  the  published experimental data upon theoretical
      models.  We would not dare to go beyond  that,  and  estimate  the
      reliability of the experimental procedure itself.
      \item
      Either  increasing  the  confidence of the existing measurement of
      the $dd \rightarrow \pi ^0 \alpha$ cross section  at  the  present
      energy  or further approaching the $\eta$ production threshold can
      improve  the  confidence  of  the  $\pi  ^0$-$\eta$  mixing  angle
      extraction. However, allowing even for the extremely high systematic
      error of the $ dd \rightarrow \pi^0 \alpha$ process (factor 5)
      the $\pi^0$-$\eta$ mixing angle can not be higher then 2.204 $^0$.
\end{itemize}

This work is supported in part by the
by the U.S.-Croatia international program under contract JF-221.
One of the authors (A.\v{S}.) is grateful for the hospitality
and the financial help of the Uppsala University,
The Swedberg Laboratory, Uppsala,
Sweden, where he has been given  a lot of useful advices and had
perfect working conditions to complete the publication in the
present form.  We are saddened to inform our collaborators that one
member of our small group, Mijo Batini\'{c}, has after a lot of suffering
deceased in 1998. 
\section*{Figure captions}
{\it Figure 1.} \\
      {Argand diagram for the $\eta\alpha$ S-wave scattering length. The
      open squares connected with the full line show the  value  of  the
      $\eta\alpha$   scattering   length   for   the  nucleon-nucleon
      correlation factor $\beta=0.28$, corresponding to the estimate  of
      reference \protect{\cite{Wil94}} for  different values of $a_{\eta
      N}$.  Full dots connected with dotted lines show the value for the
      $a_{\eta\alpha}$  for  different  $a_{\eta  N}$  values  and for
      different values of $\beta$ in steps of 0.05 fm$^{-1}$, and $0.0 <
      \beta  <  0.6$  fm$^{-1}$. The open triangle represents the $\eta
      \alpha$ S-wave scattering length value without constraints imposed
      by Equation(\protect{\ref{eq:msc}}) obtained in reference
      \vspace*{1.cm} \protect{\cite{Wil94}}.} \\
{\it Figure 2.} \\
      {The square of the absolute value of the $\eta\alpha$ scattering
      function as a function of $\eta$ cm momentum. Full circles are
      from reference \protect{\cite{Fra94}} and open circles are from
      reference
      \protect{\cite{Wil97}}.   Full, dashed and dotted lines correspond
      to the nucleon-nucleon correlation factor values $\beta$  =  0.56;
      0.42  and 0.28 fm$^{-1}$ respectively  for the $\eta N$ S-wave
      scattering value of ref.  \protect{\cite{Sva98}.   }

\end{document}